# A simple transcendental travelling wave solution and stability study for the thermophoretic motion with variable heat transmission factors on substrate-supported grapheme sheet

Yue Chan[1*], Daoju Cai[1], Kaisheng Cai[1], Shern-Long Lee[1], Rumiao Lin[1], Yong Ren[2]

[1] Institute for Advanced Study, Shenzhen University, Nanshan Distract, Shenzhen, Guangdong 518060, China
[2] Department of Mechanical, Materials and Manufacturing Enginnering, Faculty of Science and Engineering, The Unviersity of Notingham Ningbo, 199 Taikang East Road, Ningbo 315100, China

*unimelbat@hotmail.com

**Abstract**. Manually tailored wrinkled graphene sheets hold great promise in fabricating smart solid-state devices. In this paper, we employ an energy method to transform the original third-order partial differential equation (pde), i.e. Eq. (1) into the first-order pde, i.e. Eq. (8) for the thermophoretic motion of substrate-supported graphene sheets, which can be solved in terms of semi-group and transcendental solutions. Unlike soliton solutions derived using other more sophisticated techniques [9, 23], the present transcendental solution can be easily solved numerically and provides physical insights. Most importantly, we verify that the formation of various forms for wrinkling wave solutions can be determined by the evolution of equilibrium points for Eq. (1). This sheds a light on modifying the heat sources in order to control the configuration of wrinkle waves that has not been previously addressed.

## 1. Introduction

Graphene is a two-dimensional honeycomb lattice. As the thinnest and hardest nanomaterial in the earth, it has wide applications in the field of sensors, transistors, batteries, conductive polymer composites, to name just a few [1–3]. However, in reality, graphene is not a perfect 2D crystal, where local wrinkles have been observed [4, 5], which may be caused by chemical functional groups, defects and mechanical strains [6]. Wrinkles could alter its electronic conductivity [7] owing to thermal vibrations, unstable interactions and strain in 2D crystals edge instabilities [8]. In order to use the physical characteristics of wrinkles to design new nano-resonator and other functional devices, the generation and control of the wrinkle configurations is essential. Thermophoresis is a crucial phenomenon, where a single-layered graphene sheet is experiencing a force drifting from hot to cold (see Fig. 1). The thermophoretic motion of substrate-supported graphene sheets is one of the promising means to achieve the goal, where the propagation of solitons can be described by a KDV equation [9]. However, most of the existing studies focus on the propagation of such solitons [10, 11]. Few works attribute to the cause and the control of different wrinkle configurations, and such attempt is one of the major contributions for the present paper.

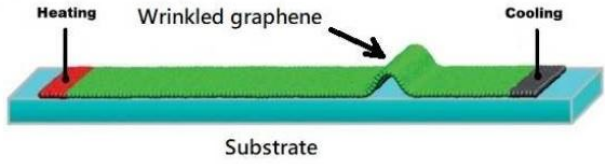

**Figure 1**: The wrinkle soliton is generated for the graphene sheets under heat transmission background.

Most of the previous theoretical works are confined to the numerical study of the wrinkling soliton dynamics, driven by the thermophoresis force using molecular dynamics simulations and first principles study [12–16]. The analytical study of soliton plays an important role in nonlinear physics. There are many techniques to construct soliton solutions of nonlinear pdes such as the Hirota's bilinear method [17, 18], Lax pair [19], the inverse scattering transformation [20, 21] and Lie groups method [22]. However, only few analytical investigations have been performed for the present topic including the construction of mixed order lump function using an extended multiple soliton method [23], the adoption of the generalized unified method to construct multiple wrinkle-like soliton solutions [10] and the bilinear transformation method [24]. Recently, numerous attempts have been tried for seeking methodologies to control nanoscale manipulation and transport [25]. However, none of the aforementioned methods are able to do so. In order to fill this gap, we employ the simple energy method [26], we are about to deduce a transcendental solution, where the factor c−f(t) (see Eq. (20)) determines the dynamics of the wrinkled graphene sheet. We further employ the stability analysis to deduce that such soliton solutions are driven by the fields, induced by the mobile equilibrium points. Hence, by scrutinizing the equilibrium points, we are able to trace out various configurations of wrinkle waves.

## 2. Theory and Various Solutions

Here, we consider a single-layered graphene, situated on a substrate, where a heat source is applied (see Fig. 1 for details). Guo et al. [9] originally observe wrinkle thermophoretic motions in substrate-supported graphene sheets from performing non-equilibrium molecular dynamics. They subsequently derive a nonlinear equation to describe such motion, which is given by $u_t + uu_x + u_{xxx} + (at + b - c_w)u_x = 0$, where a and b are system thermal conductive coefficients, and $c_w$ is a system adjusted parameter. Here, we consider a more general pde, where the linear function $at + b - c_w$ is replaced by any continuous functions f(t) with a compact support. For simply, we consider t variable in f(t) to be a mere parameter, which does not feature any wave property. Given that, the wrinkle thermophoretic motions in substrate-supported graphene sheets can be described by $u_t + uu_x + u_{xxx} + f(t)u_x = 0.$ (1)

Such nonlinear equation can be commonly solved using inverse scattering or bilinear operator transforms. However, we adopt the energy method to reduce the order of the pde so as to derive much simpler transcendental solutions. To proceed, we multiply Eq. (1) both sides by u and then integrate over x in an arbitrary spatial domain and for simplicity, we assume zero for all integration out constants to obtain $\frac{1}{2}\frac{\partial}{\partial t}||u||^2 - \frac{1}{2}\int u_x u^2 dx - \int u_x u_{xx} + f(t)\frac{u^2}{2} = 0,$ (2) where the second line of Eq. (2) is obtained using integration by parts and $||\cdot||$ is the usual $L^2$ norm in x. Due to the asymmetric nature of the integral $\int u_x u_{xx} dx$, we further simplify Eq. (2) into
$\frac{\partial}{\partial t}||u||^2 - \int u_x u^2 dx + f(t)u^2 = 0.$ (3)

Now, we base on Eq. (3) to contemplate three scenarios, namely when the $L^2$ normal of u is time independent, semi-group solution and transcendental solution.

*2.1. The norm of u is time independent*
The simplest simplification is that the norm of u is time independent so that Eq. (3) reduces to the following integral solution $u^2 = f^{-1}(t)\int u_x u^2 \, dx,$ (4) which can be termed into a transcendental solution as given below $u^2 = u_0^2 e^{\left\{\frac{u}{f(t)} - \frac{u_0}{f(0)}\right\}}$, (5)
where $u_0$ denotes the initial value of u. We note that the solution of Eq. (5) can be solved numerically at each t, where the solution can be obtained by the interception the following system of equations.
$$y = \begin{cases} u^2 \\ u_0^2 e^{\left\{\frac{u}{f(t)} - \frac{u_0}{f(0)}\right\}} \end{cases}. \qquad (6)$$

*2.2. Semi-group solution*
Now, we differentiate Eq. (3) both sides by x to get $(u^2)_t - u_x u^2 + f(t)(u^2)_x = 0$. (7)
Let $p = u_x$ and regard p as the local flux. We comment that in this stage any non-zero integration constants will be eliminated. Now let $v = u^2$, gives $v_t - pv + f(t)v_x = 0$. (8)
Multiplying both sides by the integration factor $I = e^{-\int p dt}$ to get $\{vI\}_t = -f(t)v_x I$. (9)
Let w=vI and we have $w_t + f(t)w_x = -f(t)(\int p_x dt)w$, (10)
where we can regard $p_x$ as the rate of change in local flux. Now, we change t into y to obtain
$w_y + f(t)w_x = -f(t)(\int p_x dy)w$. (11)
Using the method of characteristics, we obtain the characteristic equation as $dx/dt=f(y)$, (12)
where the characteristics line is given as $x = f(y)t + x_0$, and where $x_0$ is the initial position. Given that, Eq. (11) becomes $w_t = -f(y)(\int p_x dy)w$. (13)
We can solve Eq. (13) analytically to obtain $w = w_0 e^{-(x-x_0)(\int p_x dy)}$, (14)
where $w_0$ denotes the initial value of w and the last equality in Eq. (14) is obtained by using the characteristics line. Converting w into u and swapping y back to t, we have
$u^2 = u_0^2 e^{-\int_0^t \{(x-x_0)p_x - p\}dt'}$ and $p=u_x$. (15)
Furthermore, we will briefly discuss the existence and uniqueness of Eq. (15). For simplicity, we investigate that for Eq. (14), where $w = w_0 e^{-\int_0^t \{(x-x_0)p_x\}dt'} = S(t)w_0$. It is easy to check that S(t) satisfies the axiom of semi-group [26] if $p_x$ is transition invariant in t. Now, we remain to show that S(t), generated by the flux gradient $p_x$ is a contraction map. Suppose $x > x_0$, it is easy to check that
$|S(t)w - S(t)\hat{w}| = \left|e^{-\int_0^t \{(x-x_0)p_x\}dt'} w - e^{-\int_0^t \{(x-x_0)p_x\}dt'} \hat{w}\right| \leq |w - \hat{w}|$, (16)
provided that $\int_0^t p_x dt' \geq 0$. By Banach's fixed point theorem, the solution u, i.e. Eq. (15) exists and is unique. However, Eq. (15) has no immediate use as both p and $p_x$ are priorly determined by u. But it can be numerically solved iteratively.

*2.3. Transcendental traveling wave solution for non-equilibrium u*
In this subsection, we find that upon assuming traveling wave solution form, we can obtain the transcendental solution. Inspired by cases 2 and 3 as described above, we restate Eq. (8)
$v_t - pv + f(t)v_x = 0$, (17)
where again $p = u_x$ and $v = u^2$. Now, assuming the traveling wave solution, where $u(s) = u(x - ct)$, we obtain $-cv' - pv + f(t)v' = 0$, (18)
where ' denotes a usual derivative wrt. s. After rearranging, we obtain
$-(c - f(t))\int \frac{dv}{v} = \int p ds + c_1$, (19)
where $c_1$ is an integration constant to be determined. After some rearrangements to get the transcendental solution $u^2 = A(t)\exp(-(u/(c-f(t))))$, (20)
where $A(t)=\exp(-c_1/(-f(t)))$ and $c_1=-(c-f(0))\ln(u_0^2)-u_0$. Similar to *2.1.*, the solution of Eq. (20) can be solved numerically using $y = \begin{cases} u^2 \\ A(t)e^{-\frac{u}{c-f(t)}} \end{cases}$. (21)
Although the complexity of the exponential function in Eq. (15) has been absorbed in A(t), the solution given in Eq. (20) is obviously not unique. Physically valid solutions have to be determined.

Most importantly, we note from Eq. (20) that the term $c - f(t)$ plays a vital role in shaping the dynamics of wrinkle waves. For example, $c = f(t)$ leads to a trivial solution and $c > f(t)$ will always generate stable wave solutions, and $c < f(t)$ admits no real solutions, which are tightly related to the stability of the equilibrium points for Eq. (1). Such relations will be discussed soon in the next section. Another closed form of traveling wave soliton solutions is also explored in Appendix.

### 3. Stability Analysis

Here, we further explore the factor $c - f(t)$ by a stability analysis. We consider the following pde with a compact support. Again, we consider f(t) to be a constant for each t and not to feature any wave properties. Using a similarity solution or travel wave solution, i.e. $u = u(s) = u(x-ct)$, Eq. (1) gives

$$u''' + \frac{d}{ds}\left\{\frac{u^2}{2}\right\} + f(t)u' - cu' = 0, \text{ where } ' = d/ds. \tag{22}$$

Integrate both sides and assume compact supports for u and its derivatives to obtain

$$u'' + u^2/2 + f(t)u - cu = 0. \tag{23}$$

Upon letting $w = u'$, we can term Eq. (23) into a system of the first order odes:

$$\begin{bmatrix} u' \\ w' \end{bmatrix} = \begin{bmatrix} w := F(u, w) \\ -\frac{u^2}{2} - f(t)u + cu := G(u, w) \end{bmatrix}. \tag{24}$$

The equilibrium points of Eq. (24) are given by $(u,w)_e = (0, 0)$ or $(u,w)_e = (2c - 2f(t), 0)$. We can see that the amplitude of the wrinkle wave depends on the phase velocity c and the heat source f(t). Moreover, stable equilibrium points are dense in $R^+$ along the positive u-axis with $c > f(t)$. However, we can allow $c < f(t)$, for which (0, 0) to be stable, but the remaining condition $c \leq f(t)$ will make the equilibrium point (0, 0) unstable. In other words, the stability of equilibrium points (0, 0) and (2c − 2f(t), 0) is exclusive. Finally, all stable equilibrium points are center points.

**Proof.** The Jacobian associated with Eq. (24) is given by $J = \begin{bmatrix} F_u & F_w \\ G_u & G_w \end{bmatrix} = \begin{bmatrix} 0 & 1 \\ -u - f(t) + c & 0 \end{bmatrix}.$ (25)

**Case 1: For $(u,w)_e = (0, 0)$,** $J_{(0,0)}$ becomes $J_{(0,0)} = \begin{bmatrix} 0 & 1 \\ c - f(t) & 0 \end{bmatrix}.$ (26)

Hence the characteristic equation for Eq. (26) is given by $\lambda^2 + f(t) - c = 0$, and the associated eigenvalues are given by $\lambda_\pm = \pm(c - f(t))^{1/2}$. It is easy to observe that $c > f(t)$ results in unstable saddle points and $c < f(t)$ will give stable center points. $c = f(t)$ gives $\lambda_\pm = 0$, which is undetermined by the linear stability. Hence, we can shift our attention to Lyapunov's Theorem to test its stability. Let the test function to be $V(u,w) = u^2/2 + w^2$. It is easy to check that $V(0, 0) = 0$ and $V(u, w)$ is positive definite. However, we can find points $(\varepsilon, \eta)$, where $\varepsilon, \eta > 0$ near (0, 0) such as $V(\varepsilon, \eta)V*(\varrho, \eta) > 0$, where * denotes the derivative along the path s. It is easy to show that $V*(u,w) = uw - wu^2$. Hence, $V*(\varepsilon, \eta) = \varepsilon \eta (1- \varepsilon) > 0.$ (27)

Due also to the positive definite of V, we have $V(\varepsilon, \eta)V*(\varepsilon, \eta) > 0$ for some points $(\varepsilon, \eta)$ near the origin so that the equilibrium point is unstable when $c = f(t)$.

**Case 2 For $(u,w)_e = (2c-2f(t), 0)$,** $J_{(2c-2f(t),0)}$ becomes $J_{(2c-2f(t),0)} = \begin{bmatrix} 0 & 1 \\ f(t) - c & 0 \end{bmatrix}.$ (28)

Hence the eigenvalues $\lambda_\pm$ for Eq. (28) are given by $\lambda_\pm = \pm(f(t) - c)^{1/2}$. While $c < f(t)$ results in unstable saddle points, $c > f(t)$ constitutes stable center points. It remains to check the case of $c = f(t)$, which gives $\lambda_\pm = 0$ and $(0, 0)_e$ becomes the equilibrium point. This reduces to the first case, by Lyapunov's theorem, the equilibrium point (0, 0) is unstable. Therefore, the physically feasible solution is created when it's phase velocity exceeds the strength of the heat transmission, i.e. $c > f(t)$.

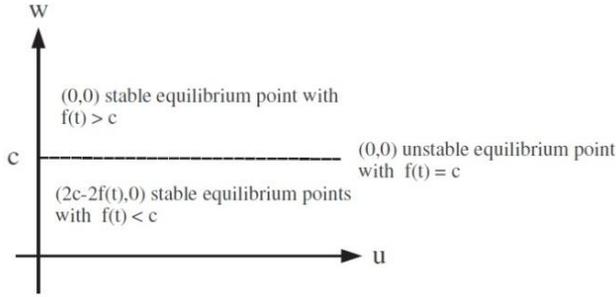

**Figure 2**: Selection rues for stability of equilibrium points for Eq. (22).

## 4. Numerical Solution and Discussion

Now, we assume the initial value of u to be 0.1, make c = 1 and use three distinct forms of f(t), i.e. periodic function 0.5 sin(t) (case a), hyperbolic functions of 0.5 tanh(t) (case b) and 0.5 sech(t) (case c) to carry out our simulations using Eq. (21), where the numerical results are given in Fig. 3.

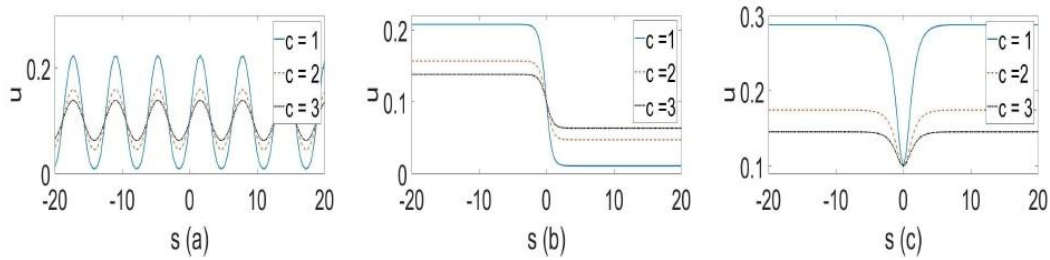

**Figure 3**: Wrinkle soliton waves for (a) f(t) = 0.5 sin(t), (b) f(t) = 0.5 tanh(t) and (c) f(t) =0.5 sech(t), respectively, and for three phase velocities, namely c = 1, 2 and 3.

As you can see in Fig. 3 that the proposed time-varying heat sources, f(t) generate snake, Z and V-like wrinkling soliton solutions for the cases (a), (b) and (c), respectively, which match well with that derived using the bilinear transformation method [24]. All the wave amplitudes satisfy the initial value, i.e. $u_0$ = 0.1 at s = 0. There exist intimate relations between u and the heat source f(t), and most crucially, the relation c > f(t) has to be satisfied for all t. We comment that during our calculation, c = f(t) results in a unique trivial solution (that is also at the characteristic line, i.e. Eq. (12) forming the forbidden regime for the propagation of solutions), and c < f(t) leads to some complex solutions as no interception of real solutions in Eq. (21) is feasible. Oscillating heat sources generate oscillating waves, while the increasing (decreasing) hyperbolic heat sources generate decreasing (increasing) wrinkle waves. Next, three different phase velocities, namely c = 1, 2 and 3 are considered, again shown in Fig. 3. An increase in phase velocities makes the oscillatory amplitude of the waves smaller due to the energy conservation. Without loss of generality, we focus on positive s and observe that an increase in phase velocities lowers the decline and enhancement of wrinkle waves in cases (b) and (c), respectively. The asymptotic solutions in cases (b) and (c) also satisfy Eq. (5), where the norm of the solution is time independent.

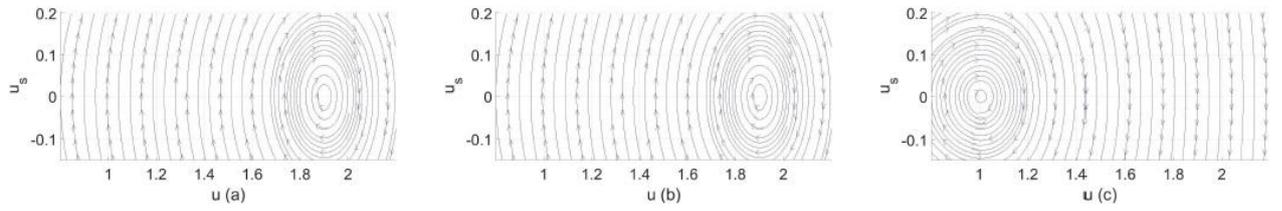

**Figure 4**: Phase diagrams at c = 1 and s = 0.1 for (a) f(t) = 0.5 sin(t), (b) f(t) = 0.5 tanh(t) and (c) f(t) = 0.5 sech(t), respectively.

In order to demonstrate the formation of miscellaneous shapes for wrinkle soliton waves and investigate the criteria of c > f(t) for generating stable soliton solutions, we turn our attention to Sec. 3, where stability analysis has been implemented. We have deduced in Sec. 3 that positive equilibrium points u = 2c − 2f(t) are always stable for c > f(t). To proceed, the phase diagrams for the three proposed cases at c = 1 and t = 0.1 are shown in Fig. 4.

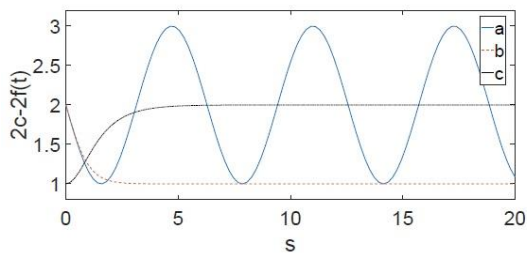

**Figure 5**: Time evolution of nonzero equilibrium points for (a) f(t) = 0.5 sin(t), (b) f(t) = 0.5 tanh(t) and (c) f(t) = 0.5 sech(t), respectively.

It clearly demonstrates all equilibrium points at u = 2 − 2f(0.1) are centers, which have been confirmed by the phase plane analysis so that any positive solution for Eq. (22) are rotating about those stable equilibrium points. Now, we investigate the time evolution of such equilibrium points, which are shown in Fig. 5. The value of equilibrium points in Fig. 4 corresponds well with that given in Fig. 5 at s = 0.1. Besides, it is intriguing to observe that the value of equilibrium points for case (a) is scillatory; decays and reaches equilibrium for case (b); but raises and reaches equilibrium for case (c) over time. Such results correlate well with that given in Fig. 3, where the initial value of u for all cases is 0.1, which is in the close vicinity of the equilibrium points and satisfies c > f(t). Therefore, we can deduce that the evolution of u is driven by the fields, generated by the moving equilibrium points over time. Conversely, we can alter the movement of equilibrium points in order to trace out various wrinkle configurations by picking correct functional forms of heat sources. Last but not least, no continuous soliton wave can be observed if c < f(t) since u = 0 is the only stable but static equilibrium point.

## 5. Conclusion

Transcendental solutions are successfully derived using the energy method, symmetry and the assumption of the compact support. Various forms of wrinkle soliton waves are predicted by the transcendental solution, which is consistent with that deduced by other more complicated methodologies such as the bilinear operator method [24]. Stability analysis for the equilibrium points of Eq. (1) is also performed, which shows that stable equilibrium points, i.e. u = 2c − 2f(t) are dense in positive real numbers and have to meet the c > f(t) criteria, and c ≤ f(t) results in no soliton wave. Lastly, we unprecedentedly show that the dynamics of wrinkle waves is driven by the fields, generated by the moving equilibrium points. This sheds a light on tracing out different wrinkle configurations by choosing the right heat sources.

**Appendix**

Here, we adopt another form of traveling wave solution, where $\eta = x - ct - \int f(t)dt$. Then, Eq. (1) after such coordinate transformation becomes $-cu_\eta + uu_\eta + u_{\eta\eta\eta} = 0$. It is easy to show that the solution of the transformed equation with the compact support is given by $u = 3c - 3c\,\tanh2\{\left(\frac{\sqrt{3c}}{2}\right)k + (\sqrt{2}/2)(x - c't)\}$, where k and $c' = c - (1/t)\int f(t)dt$ denote the integration constant and the effective speed of the traveling wave, respectively. We comment that the value of u lays between 0 and 3c and it is also interesting to observe that if we keep $c'$ constant, then f(t) will modify c and hence alter the amplitude of the soliton.